\documentclass{aa}
\usepackage{txfonts}
\usepackage{graphics,latexsym,amssymb,times,psfig}
\begin{document}
\title{The spectra of short Gamma--Ray Bursts}
\author{Giancarlo Ghirlanda \inst{1},\ \inst{2} \and  
Gabriele Ghisellini \inst{2} \and Annalisa Celotti \inst{3}}
\offprints{G. Ghirlanda; ghirlanda@merate.mi.astro.it} 
\institute{IASF, via Bassini 15, I--20133 Milano, Italy \and
Osservatorio Astronomico di Brera, via Bianchi 46, I--23807 Merate,
Italy \and SISSA/ISAS, via Beirut 2-4, I--34014 Trieste, Italy. }
%
%
\titlerunning{Spectra of short bursts}
\authorrunning{G. Ghirlanda, G. Ghisellini \& A. Celotti}
\abstract{We present the results of the spectral analysis of a sample
of short bright $\gamma$--ray bursts (GRB) detected by BATSE and
compare them with the average and time resolved spectral properties of
long bright bursts.  While the spectral parameters of short GRBs
confirm, as expected from previous works based on the hardness ratio,
that they are harder than long events, we find that this difference is
mainly due to a harder low energy spectral component present in short
bursts, rather than to a (marginally) different peak energy.
Intriguingly our analysis also reveals that the emission properties of
short GRBs are similar to the first 2 s of long events.  This might
suggest that the central engine of long and short GRBs is the same,
just working for a longer time for long GRBs.  We find that short
bursts do not obey the correlation between peak frequency and
isotropic emitted energy for any assumed redshift, while they can obey
the similar correlation between the peak frequency and isotropic
emitted luminosity.  This is consistent with (although not a proof of)
the idea that short GRBs emit a $\gamma$--ray luminosity similar to
long GRBs.  If they indeed obey the peak frequency -- isotropic
luminosity relation, we can estimate the redshift distribution of
short bursts, which turns out to be consistent with that of
long bursts just with a slightly smaller average redshift. 
\keywords{Gamma rays: bursts, observations -- X--rays: general --
Radiation mechanisms: non--thermal, thermal } }

\maketitle

\section{Introduction}
The  possible existence of  different classes  of GRBs  was considered
since  their  discovery,  and  the strongest  evidence  for  different
populations is their bimodal  duration distribution, with $\sim$1/3 of
`short' events with  a mean duration of $\sim$0.3  s, and the majority
of `long'  events with  mean duration of  $\sim$ 20 s  (Kouveliotou et
al. \cite{Kouveliotou1993}, Norris et al. \cite{Norris2000}).  Further
support to such a bimodal  behavior emerges from the analysis of their
spectral and temporal properties: short  bursts seem to be harder than
long ones  (Kouveliotou et al.  \cite{Kouveliotou1993},  Hurley et al.
\cite{Hurley1992}) and their  distributions of pulse width, separation
and number  of pulses  per bursts also  indicate that the  two classes
might be  physically distinct (Norris et  al. \cite{Norris2000}, Nakar
\& Piran  \cite{Nakar2002}).  The  distinction between short  and long
bursts has been also considered  as indication of the existence of two
distinct  progenitors.   If  the  duration  of the  GRB  emission  (as
predicted   by  the   internal   shock  model   -   see  e.g.    Piran
\cite{Piran1999} for a review) is  linked to the duration of the inner
engine  activity, short  bursts might  be  produced by  the merger  of
compact objects  (Ruffert \& Janka \cite{Ruffert1999})  while the core
collapse  of   massive  stars  would  give  raise   to  long  duration
GRBs. While the properties of long events (e.g.  redshifts, broad band
spectral  emission  and evolution,  environment  etc.,  see Hurley  et
al. \cite{Hurley  2003} for a  recent review) have been  unveiled with
increasing  details,  the  understanding  of  short  bursts  is  still
limited.  Recently, Schmidt  (\cite{Schmidt2001}) suggested that short
bursts  have  a  similar  luminosity  to  long  events.   So  far  the
characterization of  the spectral properties of  short bursts detected
by BATSE has  been based on the comparison of the  ratio of the fluxes
emitted in different (broad) energy bands (Cline et al. 1999; Yi--ping
Qui  2001).   The spectrum  of  long  GRBs,  typically represented  by
smoothly connected power laws, is different for different bursts (Band
et al.  \cite{Band1993})  and it may also considerably  evolve in time
within the  same burst  (Ford et al.   \cite{Ford1995}, Crider  et al.
\cite{Crider1997}).   This complex behaviour  compels to  consider the
complete spectrum of any GRB with high time and spectral resolution in
order  to describe  and compare  the emission  properties of  long and
short  bursts. Clearly the  main difficulty  when fitting  short burst
spectra  is  their  low signal  to  noise  ratio  due to  their  small
duration. Paciesas et  al. (\cite{Paciesas2001}) compared the spectral
parameters of short and long  bursts obtained from spectral fits: they
found that in short bursts the  low energy spectral index and the peak
energy are harder than in long events. However, the time resolution (2
s) of the spectra used to  describe the class of short bursts was much
larger  than their  typical duration  (0.3  s) and  also the  spectral
resolution  of  their  data was  low  compared  to  that of  the  data
presented  in this work.   Also the  distance scale  of short  GRBs is
still  a matter  of debate.   Their spatial  distribution seems  to be
consistent  with  low redshift  sources  (e.g.   Magliocchetti et  al.
\cite{Magliocchetti2003}, Che et  al.  \cite{Che1997}), but nothing is
directly known  due to the lack  of any redshift  measurement.  On the
other  hand, possible  correlations among  the spectral  properties of
long   bursts    have   been   recently   claimed    (Amati   et   al.
\cite{Amati2002}, see  also Yonetoku et  al.  \cite{Yonetoku2003}) and
confirmed  by  the  Hete--II   long  GRBs  and  X--Ray  Flashes  (Lamb
\cite{Lamb2003a}).   These relations  might be  key in  explaining the
still obscure  energy conversion mechanism operating in  long GRBs and
it  is thus  interesting to  investigate whether  similar correlations
also hold for short bursts.

\section{Sample selection and spectral analysis}
From                 the                 BATSE                on--line
catalog\footnote{http://cossc.gsfc.nasa.gov/cossc/batse/}    we   have
selected bright short  ($T_{90} \le 2$ s) events as  those with a peak
flux (computed on  the 64 ms timescale and  integrated over the energy
range 50--300 keV) exceeding 10 phot cm$^{-2}$ s$^{-1}$.  The hardness
ratios  of these  36 selected  short  GRBs, computed  over the  energy
ranges  1=25--50  keV, 2=50--100  keV,  3=100--300  keV, are  $\langle
HR_{21} \rangle=1.65$ and $\langle HR_{32} \rangle=5.9$.  These values
are larger than the corresponding HRs of the population of long bursts
($\langle HR_{21} \rangle=1.56$ and $\langle HR_{32} \rangle=3.8$), in
agreement  with   the  hardness--duration  relation   (Kouveliotou  et
al. \cite{Kouveliotou1993}).  The average fluences  (for energies $\ge
25$ keV)  of the selected short  GRBs is $\sim  6.2\times 10^{-6}$ erg
s$^{-1}$  (i.e.,  only  a  factor  $\sim$  2.5  lower  than  for  long
bursts). We analyzed their spectra using the Large Area Detector (LAD)
data and applied the standard spectral analysis technique (e.g. Preece
et al.   \cite{Preece2000}, hereafter P00).  Each  spectrum was fitted
over the  energy range $\sim 30$ keV  -- 1.8 MeV.  Due  to their short
duration,  the  minimum integration  time  (S/N  limited)  of the  LAD
spectra is  typically $\sim  0.2-0.4$ s, so  that we could  analyze in
most cases one  single spectrum per short GRB.  In 7  cases out of 36,
the spectrum had a low S/N over the entire energy range which resulted
in  poorly constrained  best  fit parameters.   These  cases were  not
included  in the  final  sample.  In  one  case the  analysis was  not
possible  because of  missing data.   The remaining  28  sources (with
100-300  keV  fluence $\ge  2.4\times  10^{-7}$ erg/cm$^2$),  although
belonging  to  a  complete  peak  flux limited  sample,  do  not  form
themselves a complete sample.

The spectral properties of short bursts have been compared
with the results of the spectral analysis of a sample of bright long
BATSE bursts (Ghirlanda et al. 2002 -- hereafter G02) whose spectra
(time averaged and time resolved) were fitted with different spectral
models.  The sample of G02 was selected on the basis of the burst peak
flux, similarly to the criterion applied for the sample of short
events presented in this work.

\subsection{Spectral results}

The spectrum of the selected short bursts is in most cases well fitted
by a single power law with an exponential cutoff at high energies. The
spectral parameters are the low energy power law photon index $\alpha$
and the peak energy $E_{\rm peak}$ (in keV) of the $E F_{E}$ spectrum.
This  model fit resulted  in a  lower reduced  $\chi^2$ and  in better
constrained  spectral  parameters  compared  with the  Band  function,
single  power law  and broken  power law  models.  In  most  cases the
statistics in the  high energy channels of the spectra  is too poor to
constrain the high energy power  law component of the Band model (Band
et  al.  \cite{Band1993}).   In 5  cases out  of the  28  analyzed the
reduced $\chi^{2}$ of  the powerlaw cutoff model is  high and excludes
the fit at  99\% confidence level.  In these  cases, however, the Band
model  resulted in  even higher  reduced $\chi^2$  and we  adopted the
parameters of the powerlaw cutoff  model for homogeneity with the rest
of the sample.  Tab. 1
\footnote{This is  available in electronic form at  ........  }  lists
our  selected bursts,  together with  the spectral  results  and their
errors  at  99\% confidence  level.   As  also  found by  Paciesas  et
al. (\cite{Paciesas2001}),  there is  no evidence for  any correlation
between $\alpha$, $E_{\rm peak}$ and the burst duration represented by
the $T_{90}$ parameter  given in the BATSE catalog  of Paciesas et al.
(\cite{Paciesas1999}).  The short  duration of most bursts ($T_{90}\in
[0.104,1.8]$ s) does not allow  a time resolved spectral analysis, but
in 5 cases we could extract at least two (in one case even three) time
resolved spectra  within the $T_{90}$ interval.   Nonetheless, the low
statistics   of  these   time  resolved   spectra  results   in  large
uncertainties on the spectral  parameters, with only a weak indication
of a hard--to--soft  spectral evolution similar to what  found in long
GRBs  (e.g. Ford et  al.  \cite{Ford1995}).   This supports  the trend
found  from   the  analysis  of  the  time   resolved  hardness  ratio
(e.g. Cline et al. \cite{Cline1999}).

In 6 short  GRBs the low energy spectrum is  harder than the optically
thin  synchrotron limit $\alpha\sim  -2/3$ (Katz  1994).  In  one case
$\alpha>0$.  A similar fraction  of long bursts violating these limits
has  been  found  in  the  population  of long  GRBs  (Crider  et  al.
\cite{Crider1997}, P00, Ghirlanda et al. \cite{Ghirlanda2003}).

\section{Short vs Long GRBs}

We compared  the spectral properties of  our short GRBs  with those of
the  long--bright events  of  G02. For  homogenity  we considered  the
spectral parameters  of long bright  bursts obtained from the  fits of
the same model, i.e.  a cutoff--powerlaw, adopted for the short ones.

\begin{figure}[hbt]
\begin{center}
\resizebox{\hsize}{!}{\includegraphics[82pt,7pt][472pt,415pt]{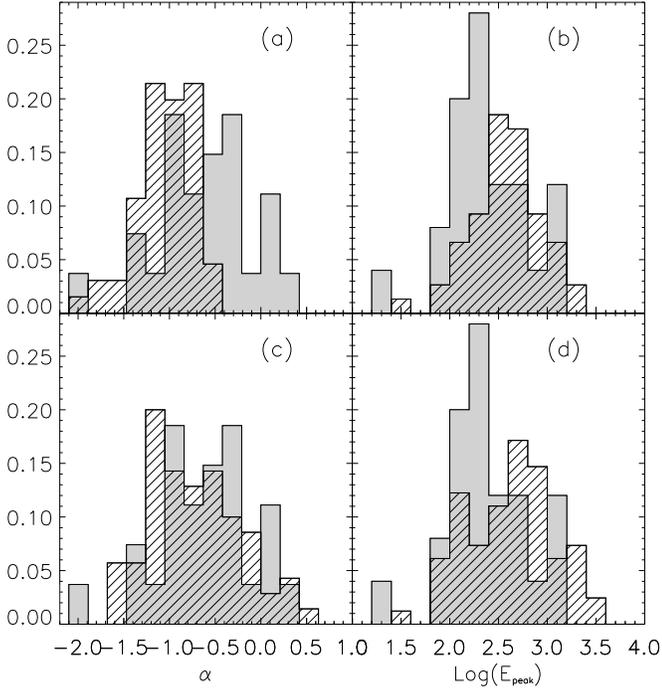}}
\vskip 0.2 true cm
\caption[]{ Distributions  of the  spectral parameters of  long bursts
({\it hatched histogram})  of the G02 sample compared  to short bursts
({\it  shaded  histogram}).  Top  panels: low  energy  spectral  index
$\alpha$ (a) and peak energy  of the $EF_{E}$ spectrum (b) considering
the  time   integrated  spectrum  of  long   bursts.   Bottom  panels:
comparison of the same spectral parameters for the first 2 sec of long
bursts. The distributions are normalized to their total number.  }
\label{fig1}\end{center}
\end{figure}

Firstly  we  compare the  time  integrated  spectral parameters.   The
 distributions  of $\alpha$  and  $E_{\rm peak}$  for  long and  short
 bursts are  reported in Fig.   \ref{fig1} (top panels).   Long bursts
 have an average peak  energy ($\langle E_{\rm peak}\rangle=520\pm 90$
 keV)   slightly  larger   than   that  of   short  events   ($\langle
 E_{\rm peak}\rangle=355\pm 30$ keV).  The distribution of the latter ones
 spreads between a few keV  and a few MeV.  A Kolmogorov--Smirnov (KS)
 test  gives a  probability of  $\sim$0.8\% that  the two  samples are
 drawn from the same  parent population.  A more significant diversity
 results  from the comparison  of the  distributions of  $\alpha$: the
 average  values  are   $\langle  \alpha  \rangle=-1.05\pm  0.14$  and
 $\langle  \alpha  \rangle=-0.58\pm 0.10$  for  long  and short  GRBs,
 respectively (KS probability of 0.04\%).
%
%
Then we  conclude that the hardness--duration  relation, discussed for
the BATSE bursts (Kouveliotou  et al. \cite{Kouveliotou1993}, Cline et
al.  1999, Yi-ping Qui 2001), is  caused more by short bursts having a
harder low  energy spectral  slope rather than  a higher  peak energy.
Indeed,  if  $E_{\rm peak}$  was  the only  parameter  characterizing  the
spectral hardness, short events would  be softer than long ones.  This
conclusion is further supported  by the search of correlations between
the hardness ratios  and the spectral parameters of  our short bursts.
The only significant trend is  of harder $\alpha$ for larger HR$_{32}$
(Spearman's correlation coefficient  $r\simeq0.6$ with null hypothesis
probability  of $10^{-3}$).   A  larger sample  of  bright BATSE  long
bursts was  studied by  Preece et al.   (\cite{Preece2000}).  Although
their sample  has not  been fitted homogeneously  with the  same model
their spectral results also confirm  our finding that short bursts are
harder than long GRBs because of a harder index $\alpha$ rather than a
higher $E_{\rm peak}$.

The  average  spectral  parameters  give  only an  indication  of  the
spectrum,  which is  likely to  evolve in  time, in  all  its spectral
parameters   (e.g.,   $\alpha$,  $E_{\rm   peak}$   etc.,  Crider   et
al. \cite{Crider1997}).   Therefore, in order to  test the tantalizing
hypothesis that the spectrum of short bursts is similar to the initial
emission  phases  of long  events,  we  considered  the time  resolved
spectral parameters reported for the  G02 sample relative to the first
seconds since the burst onset.

The spectrum of short bursts, as described by $\alpha$ and $E_{\rm peak}$,
is more similar to the first  2 s than the integrated spectrum of long
events, as  shown in Fig.~1  (bottom panels): the low  energy spectral
index distributions are similar (with  a KS probability of 83\%) while
the  peak energy distributions  still indicate  that short  events are
softer than long ones (with a KS probability of 10\%).

Another appealing possibility is that short bursts might be similar to
the peak spectra  of long bursts (i.e. they represent  the `tip of the
iceberg' in the  long GRB light curve). To this  aim we extracted from
the catalog  of Preece et  al.  (2000) only those  spectra accumulated
around the  peak with  an integration time  (centered around  the peak
time) at  least as long as  the average duration of  short bursts (0.3
s). From  the comparison of  the spectral parameters  distributions we
conclude that short  bursts are still different --  especially for the
$E_{\rm peak}$ value  -- from the peak  of long GRBs as  also indicated by
the  small  KS probability  (0.4\%).  The  low  energy spectral  index
instead, similarly to what found  from the comparison with the first 2
seconds of  long events, presents  a distribution similar to  that for
the  peak spectra  of long  events (with  a KS  probability  of 23\%).
Moreover,  we stress that  if the  fits were  performed with  the Band
model (regardless  of the indetermination of the  high energy spectral
slope) we  should find  a systematically lower  peak energy  than what
found   with  the  powerlaw   cutoff  model   (e.g.   Preece   et  al.
\cite{Preece2000}, Ghirlanda et al.  \cite{Ghirlanda2002}). This would
strengthen our conclusions.

\section{The energy and luminosity of short bursts}

\begin{figure}[hbt]\begin{center}
\resizebox{\hsize}{!}{\includegraphics[103pt,372pt][535pt,696pt]{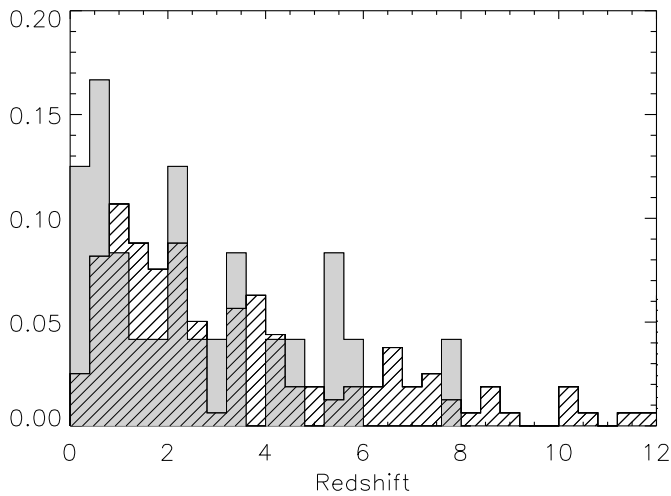}}
\caption[]{Redshift distributions for the sample of short bursts ({\it
shaded histogram})  inferred from the $L_{\rm iso}$  vs $E_{\rm peak}$
correlation.  For  comparison we report  also the $z$  distribution of
long   bursts   ({\it    hatched   histogram})   obtained   from   the
variability--luminosity   correlation   (Fenimore   \&   Ramirez--Ruiz
\cite{Fenimore2000}).}
\label{fig3}\end{center}
\end{figure}

To  further test  the relationship  between  short and  long GRBs,  we
considered  the recently  found  spectral correlations  for long  GRBs
between  $E_{\rm peak}$  and the  equivalent isotropic  energy $E_{\rm
iso}$ in  the source reference frame (Amati  et al. \cite{Amati2002}).
A similar  result was found  from a sample  of BATSE bursts  (Lloyd et
al. \cite{Lloyd2000}, Bloom et  al.  \cite{Bloom2001}) and also by the
Hete--II  long  bursts,  with  the  inclusion  of  2  X--Ray  Flashes,
extending    this    relation     to    low    energies    (Lamb    et
al. \cite{Lamb2003a}). Moreover  Yonetoku et al. (\cite{Yonetoku2003})
found a  similar correlation between the  isotropic luminosity $L_{\rm
iso}$ and $E_{\rm  peak}$ of the BATSE and  BeppoSAX samples. If short
GRBs  are  similar  to  long  events  they  might  satisfy  the  above
correlations.  However, the redshift of short bursts is unknown.  Thus
we can only  verify whether the observed spectral  properties of short
bursts, scaled in the source  rest frame (for any $z$), are consistent
with the above  correlations.  In other words, from  the spectral fits
of  short bursts  we can  derive the  peak energy  ($E_{\rm peak}^{\rm
obs}$) and the  fluence ($F^{\rm obs}$) integrated over  the same band
used by  Amati et  al.  (\cite{Amati2002}) and  Bloom et  al.  (2001),
i.e.  1 keV -- 10 MeV.  These can be converted in the source reference
frame   for    any   redshift:   $E_{\rm    iso}(z)=F^{\rm   obs}(4\pi
D_{L}^{2})/(1+z)$,  $E_{\rm peak}(z)=E_{\rm peak}^{\rm  obs}(1+z)$ and
$L_{\rm  iso}(z)=F^{\rm  obs}\times(4\pi  D_{\rm L}^{2})/T_{90}$  [for
cosmological      parameters      $(H_{0},\Omega_{\Lambda},\Omega_{\rm
m})=(65,0.7,0.3)$].    These   were   compared   with   the   relation
$E_{\rm iso}\propto   E_{\rm   peak}^{1.93}$  found   by   Amati  et   al.
(\cite{Amati2002}) for  a redshift range 0.001-10.  The  same was done
for the luminosity  $L_{\rm iso}(z)$.  We find that  for any given $z$
(up to  10) short bursts (except  for 2 cases) populate  a region {\it
below}  the $E_{\rm  iso}$--$E_{\rm peak}$  correlation of  long GRBs.
The luminosity of short events
\footnote{ Note that  $L_{\rm iso}$ for short bursts  is obtained from
the  spectrum  integrated   over  their  typical  duration  ($\sim$0.5
sec). Similarly Yonetoku et al. (2003) derive this luminosity for long
bursts either  from the spectrum integrated  around the GRB  peak on a
comparable timescale  (for BATSE  GRBs) or by  dividing for  the burst
duration  (for   SAX  GRBs).   }   is  instead   consistent  with  the
correlation     proposed    for     long    bursts     (Yonetoku    et
al. \cite{Yonetoku2003}).   Following the spectral  results suggesting
that short GRB might be similar to  the first $\sim 1$ s of long ones,
we scaled the $L_{\rm  iso}$--$E_{\rm peak}$ correlation (computed for
the peak spectra of long events) to the first second of their emission
by considering  the typical ratios  between fluxes and  $E_{\rm peak}$
estimated at the peak and those integrated over the first second.  The
curves of $L_{\rm  iso}(z)$ of short bursts are  still consistent with
this  relation.  Although  neither  obvious nor  unique to  interpret,
these results  are at  least consistent with  the hypothesis  that the
luminosity of short  GRBs is comparable to that  of long bursts within
the first  seconds, with instead a lower  isotropic equivalent energy.
In this case  the different duration might be  responsible for a lower
total energy emitted in short bursts. Under such hypothesis it is also
possible to tentatively infer the redshift distribution of (our) short
bursts by assuming that they satisfy the $L_{\rm iso}$--$E_{\rm peak}$
correlation.   The  resulting $z$  distribution  (Fig. \ref{fig3})  is
compared to that of a sample of long bursts for which the redshift has
been  inferred from  the claimed  correlation between  the variability
properties of  their prompt emission light curve  and their luminosity
(Fenimore \& Ramirez--Ruiz 2001). Short bursts have a slightly smaller
average redshift compared to the sample of long events.

%
%

We stress  that this  result is heavily  dependent on  the assumptions
that short GRBs i) have the same luminosity of long ones, and ii) obey
the same  $L_{\rm iso}$--$E_{\rm  peak}$ correlation followed  by long
GRBs. Both these  assumptions will be tested when  a reasonable number
of redshift of short GBRs will be known.

\section{Conclusions}

We selected a sample of short bright bursts from the BATSE catalog and
performed a standard spectral  analysis in order to characterize their
spectrum. No correlation was found between the spectral parameters and
the global  properties (duration and  flux) of these bursts.   The low
energy spectral  index $\alpha$ is  distributed between --2  and small
positive values and  the peak energy $E_{\rm peak}$  is between 20 keV
and 2 MeV.   Similarly to long bursts (e.g. Preece  et al. 2000), some
short bursts have a low energy spectrum harder than the optically thin
synchrotron limit.

The comparison of the spectral  properties of short and long GRBs (G02
sample) revealed  that: {\it  (i)} the higher  hardness of  short GRBs
with respect to  long events (typically described in  terms of fluence
hardness ratio) is  the effect of a harder  low energy spectrum rather
than of  a marginally different  peak energy; {\it (ii)}  the spectral
properties of  short bursts are similar to  those of the first  2 s of
long GRBs.

Short bursts  are then harder  than the time--average spectra  of long
GRBs,  but their properties  are compatible  with a  similar mechanism
operating  at the  beginning  of all  bursts,  independently of  their
duration.

Short bursts cannot obey the energy--peak frequency correlation found
for long bursts (Amati et al.  \cite{Amati2002}, Lamb et al.
\cite{Lamb2003a}), but they {\it could} obey the similar correlation
found by Yonetoku et al. (\cite{Yonetoku2003}) between the luminosity
and the peak frequency.  This may suggest that short bursts have the
same luminosity, but lower total energy, than long bursts.  {\it If}
this is the case the redshift distribution inferred for the short
bursts of our sample is similar to that of long events, with a
slightly smaller average redshift.  

\begin{acknowledgements}
This research  has made use of  data obtained through  the High Energy
Astrophysics Science Archive  Research Center Online Service, provided
by the NASA/Goddard  Space Flight Center. We thank F. Tavecchio 
for discussion
and the referee for useful comments.
\end{acknowledgements}

\newpage 

\thispagestyle{empty}

\setcounter{table}{0}
\begin{table*}[hbt]
\caption[]{Spectral results for the sample of bright short bursts.}
\begin{tabular}{cccccccrr}
\hline
GRB & TR \#$\ ^{\mathrm{a}}$	&	$t_{90}$ $^{\mathrm{b}}$    &	  $P_{64ms}$ $^{\mathrm{c}}$       &	  $\alpha$    &	 $E_{peak}$    &	$\chi^{2}/dof$   &	$F^{\mathrm{d}}\ \ \ \ \ \ \ $ & Fluence$^{\mathrm{h}}$\ \ \ \  \\
    &	&	sec	    &	phot/cm$^{2}$ sec   &                 &  keV	       &			&      erg/cm$^2$ sec & erg/cm$^2$\ \ \ \   \\
\hline \noalign{\smallskip} 
910609	&	298	&	 0.455$\pm$0.065	    &    56.1$\pm$1.2       &  -0.5$\pm$0.8   &    122$\pm$53       &     113.5/102      &          (2.6$\pm$0.1)E-07    &  (1.82$\pm$0.07)E-07 \\
910626	&	444	&	 0.256$\pm$0.091	    &    28.6$\pm$0.7	    &  -0.8$\pm$0.2   &    128$\pm$44	    &	  115.4/102	 &          (1.9$\pm$0.03)E-06   &  (4.8$\pm$0.07)E-07 \\
911119	&	1088	&	 0.192$\pm$0.091	    &	 11.9$\pm$0.6	    &  0.1$\pm$1.0    &    143$\pm$110	    &	  123.3/104	 &          (7.1$\pm$0.3)E-08   &  (7.3$\pm$0.3)E-08 \\
920229	&	1453	&	 0.192$\pm$0.453	    &	 11.9$\pm$0.6	    &  -0.15$\pm$0.08 &	   173.8$\pm$89	    &	  87.6/108	 &	    (2.5$\pm$0.1)E-07   &  (1.76$\pm$0.07)E-07 \\
920414	&	1553	&	 0.96$\pm$0.143	            &	 13.7$\pm$0.5	    &  -0.5$\pm$0.3   &	   548$\pm$310      &	  108/95	 &	    (8.4$\pm$0.2)E-06   &  (8.6$\pm$0.2)E-06 \\
921123	&	2068	&	 0.591$\pm$0.06	            &	 15.6$\pm$0.6 	    &  -0.2$\pm$0.3   &    172$\pm$50	    &	  129/107	 & 	    (1.0$\pm$0.03)E-06   &  (0.38$\pm$0.01)E-06 \\
930110	&	2125	&	 0.223$\pm$0.013	    &	 15$\pm$0.5	    &  -0.5$\pm$0.2   &	   583$\pm$150	    &	  86/102	 &	    (5.2$\pm$0.07)E-06   &  (1.66$\pm$0.02)E-06 \\
930329	&	2273	&	 0.224$\pm$0.066	    &	 18.5$\pm$0.6 	    &  -0.18$\pm$0.21 &	   290$\pm$123	    &	  88/99		 &	    (8.5$\pm$0.4)E-07   &  (3.8$\pm$0.17)E-07 \\
930428	&	2320	&	 0.608$\pm$0.041	    &	 11$\pm$0.5	    &  -0.6$\pm$0.1   &    184$\pm$45	    &     82/103	 &	    (1.9$\pm$0.1)E-06   &  (1.8$\pm$0.1)E-06 \\
930905	&	2514	&	 0.2$\pm$0.094	            &	 28$\pm$0.7	    &	-0.8$\pm$0.1  &	   194$\pm$38	    &	  112/100	 &	    (4.3$\pm$0.1)E-06   &  (1.1$\pm$0.02)E-06 \\
931101	&	2614	&	 0.296$\pm$0.057	    &	 10$\pm$0.5	    &	-1.0$\pm$0.1  &    163$\pm$240	    &	  90/108	 &	    (4.6$\pm$0.1)E-06   &  (1.47$\pm$0.03)E-06 \\
931205	&	2679	&	 0.256$\pm$0.091 	    &	 13.7$\pm$0.5	    &   -0.3$\pm$0.2  &	   1025$\pm$343     &	  145/106	 &	    (1.1$\pm$0.07)E-05   &  (0.35$\pm$0.02)E-05 \\
931229	&	2715	&	 0.384$\pm$0.091 	    &	 10.4$\pm$0.5	    &	0.14$\pm$0.11 &    1031$\pm$319	    &	  111/107	 &	    (2.7$\pm$0.4)E-05   &  (0.9$\pm$0.1)E-05 \\
940329	&	2896	&	 0.456$\pm$0.033	    &	 10.4$\pm$0.4	    &   -0.8$\pm$0.2  &	   90$\pm$29	    &     113/106	 &	    (1.8$\pm$0.05)E-06   &  (0.57$\pm$0.01)E-06 \\
940415	&	2933	&	 0.32$\pm$0.091             &	 10.7$\pm$0.4	    & 	0.2$\pm$0.6   &	   232$\pm$148	    &	  153/107	 & 	    (5.9$\pm$0.3)E-07   &  (3.4$\pm$0.2)E-07 \\
940717	&	3087	&	 1.152$\pm$0.091	    &	 18.6$\pm$0.5	    &   -1.1$\pm$0.1  &	   204$\pm$55	    &	  139/107	 &	    (3.4$\pm$0.1)E-06   &  (4.6$\pm$0.1)E-06 \\
940902	&	3152	&	 1.793$\pm$0.066	    &	 25.3$\pm$0.7	    &   -0.2$\pm$0.3  &    937$\pm$265      &     129/107	 &	    (2.4$\pm$0.07)E-05   &  (3.6$\pm$0.1)E-05 \\
940918	&	3173	&	 0.208$\pm$0.025            &	 14.9$\pm$0.6	    &   -1.0$\pm$0.1  &	   561.8$\pm$300    &	  142/105	 & 	    (1.8$\pm$1)E-06   &  (0.6$\pm$0.3)E-06 \\
950211	&	3412	&	 0.068$\pm$0.006	    &	 54.8$\pm$0.7	    &   -1.3$\pm$0.5  &    75.5$\pm$60	    &	  92/103	 &	    (8.9$\pm$1.5)E-07   &  (2.6$\pm$0.4)E-07 \\
960803	&	5561	&	 0.104$\pm$0.011	    &	 19.3$\pm$0.4	    &   -2.7$\pm$0.2  &	   $\ge$ 28	    &	  108/109	 &	    (1.2$\pm$0.5)E-06   &  (1.0$\pm$0.4)E-06 \\
970315	&	6123	&	 0.186$\pm$0.042 	    &	 12.8$\pm$0.4	    &   -0.2$\pm$1.0  &    135$\pm$100	    &     119/108	 & 	    (9.9$\pm$0.1)E-08   &  (0.85$\pm$0.01)E-07 \\
970704	&	6293	&	 0.192$\pm$0.091	    &	 88.5$\pm$1.0	    &   -1.2$\pm$0.02 &    $\le$ 1800       & 	  132/109	 &	    (1.1$\pm$0.2)E-04   &  (0.25$\pm$0.04)E-04 \\
971218	&	6535	&	 1.664$\pm$0.143	    &	 11.8$\pm$0.3	    &   -0.9$\pm$0.08 &	   1202$\pm$407	    &     150/108	 & 	    (6.1$\pm$0.2)E-06   &  (7.2$\pm$0.2)E-06 \\
980310	&	6635	&	 1.152$\pm$0.143	    &	 12$\pm$0.3	    &	-1.9$\pm$0.1  &	   20$\pm$15	    &	  109/108	 &	    (2.0$\pm$0.01)E-06   &  (4.42$\pm$0.02)E-06 \\
980330	&	6668	&	 0.116$\pm$0.006            &	 39$\pm$0.6	    &   -0.3$\pm$0.4  &	   204$\pm$118      &	  126/107	 &	    (8.2$\pm$0.4)E-07   &  (5.0$\pm$0.2)E-07 \\
981226	&	7281	&	 1.664$\pm$0.143	    &	 16.8$\pm$0.4	    &   -0.8$\pm$0.1  &	   148$\pm$28       &	  138.6/107	 &	    (9.1$\pm$1.0)E-07   &  (7.8$\pm$0.8)E-07 \\
991002	&	7784	&	 1.9$\pm$0.5                &	 10.2$\pm$0.3	    &   -0.8$\pm$0.3  &	   164$\pm$80	    &     154/108	 & 	    (4.1$\pm$0.2)E-07   &  (5.6$\pm$0.2)E-07 \\
000108	&	7939	&	 1.039$\pm$0.072            &	 10.7$\pm$0.3	    &	-0.0$\pm$0.2  &    128$\pm$90	    &     143/107	 & 	    (2.2$\pm$0.05)E-06   &  (0.7$\pm$0.01)E-06 \\
\hline
\end{tabular}
\begin{list}{}{}
\item[$^{\mathrm{a}}$] GRB trigger number (BATSE catalog).
\item[$^{\mathrm{b}}$] GRB duration (BATSE catalog).
\item[$^{\mathrm{c}}$] Integrated flux in the energy range 50 keV -
300 keV.
\item[$^{\mathrm{d}}$] Observed energy flux from the
best fit model over the energy range 1 keV -- 10
MeV.
\item[$^{\mathrm{h}}$] Fluence in the range 50-300 keV.
\end{list}
\end{table*}


\end{document}